\documentclass{article}
\pdfoutput=1
\usepackage{jcappub}

\def\lbldef#1#2{\expandafter\gdef\csname #1\endcsname {#2}}
\def\eqn#1#2{\lbldef{#1}{(\ref{#1})}
\begin{equation} #2 \label{#1} \end{equation}}

\def\href#1#2{#2}
                                                                              
\def\tvphi{{q}}

\title{The effective Lagrangian of dark energy from observations}

\author[a]{Raul Jimenez,}
\author[b]{P. Talavera,} 
\author[a]{Licia Verde,}
\author[c]{Michele Moresco,}
\author[c]{Andrea Cimatti,}
\author[d]{and Lucia Pozzetti}

\affiliation[a]{ICREA \& ICC, Institut de Ciencies del Cosmos, Universitat de Barcelona (IEEC-UB), Marti i Franques 1, Barcelona 08028, Spain}
\affiliation[b]{DFEN \& ICC, Universitat Polit\`ecnica de Catalunya, Comte Urgell 187, Barcelona, Spain}
\affiliation[c]{Dipartimento di Astronomia, Universita di Bologna, via Ranzani 1, 40127 Bologna, Italy}
\affiliation[d]{INAF - Osservatorio Astronomico di Bologna, via Ranzani 1, 40127 Bologna, Italy}

\emailAdd{raul.jimenez@icc.ub.edu}
\emailAdd{pere.talavera@icc.ub.edu}
\emailAdd{liciaverde@icc.ub.edu}
\emailAdd{michele.moresco@unibo.it}
\emailAdd{a.cimatti@unibo.it}
\emailAdd{lucia.pozzetti@oabo.inaf.it}

\abstract{Using observational data on the expansion rate of the universe ($H(z)$) we constrain the effective Lagrangian of the current accelerated expansion. Our results show that the effective potential is consistent with being flat  i.e., a cosmological constant; it is also consistent with the field moving along an almost flat potential like a pseudo-Goldstone boson. We show that the potential of dark energy does not deviate from a constant  at more than $6$\% over the redshift range $0 < z < 1$. The data can be described by just a constant term in the Lagrangian and do not require any extra parameters; therefore there is no evidence for  augmenting the number of parameters of the LCDM paradigm. We also find that the data justify the effective theory approach to describe accelerated expansion and that the allowed parameters range satisfy the expected hierarchy. Future data, both from cosmic chronometers and baryonic acoustic oscillations, that can measure $H(z)$ at the \% level, could greatly improve constraints on the flatness of the potential or shed some light on possible mechanisms driving the accelerated expansion. Besides the above result, it is shown that the effective Lagrangian of accelerated expansion can be constrained from cosmological observations in a model-independent way and that direct measurements of the  expansion rate $H(z)$ are  most useful to do so.}

\begin{document}

\maketitle

\section{Introduction}

The Universe is presently accelerating \cite{SN,LSS,Stern,Jimenez} and the best fit model to current data predicts that  it will eventually enter a de Sitter phase.  However we have no satisfactory theoretical explanation of what is driving this phase of accelerated expansion and/or  the physical mechanism behind it. One approach to understand the physical mechanism behind expansion is to construct effective theories that then limit the, in principle, infinite choices for the Lagrangian to just a few parameters \cite{Weinberg:2008hq,Creminelli,JTV11,Bloomfield:2011wa}.

The effective Lagrangian approach provides, in our opinion, a model independent route to constrain the nature of expansion. Any other approach will necessarily invoke extra assumptions about the nature of expansion in order to reduce the number of infinite parameters to a handful of measurable ones.  For example, the most widely used approach is to assume that
the expansion is driven by a scalar field with constant equation of state $p
= w \rho$ and use observations to find $w$ ($p, \rho$ are the pressure and
energy density of the scalar field). Of course, in this framework, a
cosmological constant corresponds to $w=-1$ while scalar fields with
dynamics deviate from this value. It is clear that one cannot assume $w$ to
be constant with time, and that a more general scenario would require a
reconstruction of $w(t)$. It is important to keep in mind that this description is a drastic simplification as it  does not cover all possible scenarios (see e.g., Ref.~\cite{Simon}), and  there could be  models where $p, \rho$ are not even defined,  but from observations of any expansion history  one could always reconstruct  an effective $w(t)$. In this case  $w(t)$ would have no physical meaning in terms of properties of a fluid or a scalar field and would be of difficult or ambiguous  theoretical interpretation. Even with this drastic simplification, it is very challenging
observationally to determine $w(t)$  e.g., Ref.~\cite{Simon}. The reason of this difficulty is that, unless one measures directly the first derivative of the Hubble parameter $H(t)$, one can always trade variations at one redshift for another, thus leaving the determination of the dark energy potential  poorly unconstrained.  

Another approach is to build theoretical models of what may be driving the expansion in order to then constrain them with observations. However, the number of models is ever growing and there is no clear guidance as to what models, or family of models, should be the ones that are physically justified  e.g.,\cite{Peebles:2002gy,Copeland:2006wr}. 

We have recently constructed an effective theory of expansion \cite{JTV11} that self-consistently provides a natural cut-off to the effective theory and therefore the number of free parameters in the Lagrangian is not arbitrarily chosen. This has provided us with a  general Lagrangian of the accelerated expansion with a finite number of free parameters (being the leading ones) to be determined from observations.  In this paper we set up to do this using the most recent determination of  $H(z)$  \cite{Moresco2012}. The impatient, or more observationally oriented, reader can go directly to Fig.~\ref{fig:3} which shows the main result of this work. The rest of the paper explains and motivates this finding from a theoretical point of view. 

\section{Theoretical set-up}

In Ref.~\cite{JTV11} it was presented an effective approach to describe accelerated expansion. We start here by reviewing it and then will re-cast it in terms more suitable for data analysis. Due to the lack of knowledge on the spectrum and symmetries of the expansion field, our approach is based on one single assumption:  scalar particles are the main component of dark energy, and on two considerations: {\sl i)}  The accelerated expansion is a relatively recent phenomena in the history of the Universe; our aim therefore  is to capture only the most recent features of our expanding universe thus {\sl all} the high-energy effects are integrated out and their effect are collected in a few low-energy constants.  {\sl ii)} Although it is well known   that a  cosmological constant provides a good description of the current data, instead of imposing this condition, we take this to be the limiting case of a scalar field in a very shallow potential. This field has its natural dimensions, $[{\rm scalar}]={\cal O}(\mu)$,
representing $\mu$ the energy scale characterizing the phenomena which we have integrated out to obtain our low-energy description. Naively one would expect that the scale $\mu$ is identified with the Planck scale $m_{\rm p}$. This would be indeed the case if at next-to-leading order all the diagrams with external scalars will be mediated only by hard gravitons, see below. As we ignore the underlaying mechanism for symmetry breaking, we have no prior knowledge for the value of this scale. However, as we shall see below, some educated guess can be deduced by inspecting the fits to the data, see Fig. \ref{fig:3}. 

The dark energy potential is considered to be smooth and we assume that deviations from a pure constant are due to small variations of the field. Concerning the treatment of gravity the reasoning goes as follows: our space-time is almost flat today, thus the dynamics of the scalar field must be,
at leading order, insensitive to higher order gravity invariants. In addition General Relativity fulfills precision tests on a wide range of scales. This motivates our choice for the power counting: by construction the effects of the scalar field is sub-leading to the gravitational one.
We treat the gravity sector in a perturbative fashion. Because the spatially flat Robertson--Walker metric is conformally flat it has a vanishing Weyl tensor, and so the Weyl tensor starts with a term of first order in perturbations. For this reason we use the Weyl tensor instead of  $R^2$ and $R_{\mu\nu}R^{\mu\nu}$ as in 
Ref.~\cite{Weinberg:2008hq}.

Collecting the first terms in the expansion one gets
\begin{eqnarray}
\label{termsm6}
{\cal S}   = \int d^4\,x  \sqrt{ -g} \left( R (m_{\rm p}^2+ \alpha_1 \varphi+\alpha_2 \varphi^2) + 
f_9  C_{\mu\nu\alpha\beta}C^{\mu\nu\alpha\beta} - {\frac{1}{2}} g^{\mu\nu} \varphi_{\,,\mu} \varphi_{\,,\nu} -V(\varphi) \right)\,,
\end{eqnarray}
with the effective Newton's constant given by 
\begin{equation}
\frac{1}{ 8 \pi G_{\rm eff}}= m_{\rm p}^2+ \alpha_1 \varphi+\alpha_2 \varphi^2\,.
\end{equation}
Note that $\alpha_1$ and $\alpha_2$ are perturbatively small.
After a conformal transformation the leading order action takes the form
\begin{eqnarray}
\label{born4}
 { {\cal S}} = \int d^4\,x \sqrt{- {g}} \left( {\frac{m_{\rm p}^2}{2}}  {R}  
+ f_9  {C}_{\mu\nu\alpha\beta} {C}^{\mu\nu\alpha\beta}-  {\frac{1}{2}}{g}^{\mu\nu}  {\partial}_\mu \tvphi  {\partial}_\nu \tvphi - U(\tvphi)\right)\,,
\end{eqnarray}
with 
\begin{equation}
\label{poteff}
U(\tvphi) = {\lambda_0+\lambda_1 q + \lambda_2 q^2 + \lambda_3 q^3 + \lambda_4 q^4}+\ldots\,.
\end{equation}
Here $q$ is the field responsible for the expansion and $f_9, \lambda_{1-4}$ are the free parameters to be determined by experiments to  constrain the theory.  Higher orders, $\lambda_i (i\ge 5)$, are suppressed. Although the potential can be obtained naively with just pure dimensional analysis it expression matches with the polynomial part of the one loop diagrams involving scalars in external legs with scalars and {\sl hard} gravitons in internal propagators once the latter have been integrated out. As by product 
of this partial integration an scale $\mu$, that can be lower than the Planck scale, appears.
We look for approximate global symmetries $\varphi\to\varphi + {\rm const.}$ Because string theory does not respect global symmetries \cite{Banks:1988yz}, taking this option we assume that at this stage quantum-gravity effects play no dominant role and the physical effect is purely classical. Henceforth we explicitly impose $q=0$ at  the present day  and  allow a constant term, $\lambda_0$, in the potential $U(q)$ as a reference to the LCDM model, but that it can take zero value if the data prefer to do so. In that manner the Universe is spatially flat i.e.  $\Omega_0+\lambda_0=1$.

The Friedman's and the resulting field equations for the action presented in (\ref{born4})
are akin to a scalar-tensor cosmology 
\eqn{friedman1}{
H^2 = \left({\dot{a}\over a}\right)^2= {1\over 3} \left( \rho_m + \rho_q\right)\,,\quad
{ \ddot{a}\over a}= {1\over 6} (\rho_m+3 p_m+\rho_q+3p_q)\,,
}
and
\eqn{eomsca}{
\ddot{\tvphi} +3 H \dot{\tvphi}-U^\prime=0\,.
}
Dotted quantities stand for derivatives w.r.t. time and prime quantities denote derivatives w.r.t.  the field $q$; $\rho_m$ denotes the matter density parameter and we assume that the Universe is dominated by  collisionless, pressureless matter,  $p_m=0$. 

Given a set of parameters $\lambda_i$, note that the first Friedman equation and the Klein Gordon  equation can be rewritten in terms of the redshift using the fact that $H(z)=-1/(1+z)dz/dt$  and then can be combined in a single differential equation for the field as a function of redshift $q(z)$:
\begin{equation}
\frac{1}{6}\left(\frac{dq}{dz}\right)^2(1+z)^2U'-\left[U(1+z)+\rho_{m,0}(1+z)^4\right]\frac{dq}{dz}-U'=0,
\label{eq:master}
\end{equation}
where $\rho_{m,0}$ denotes the present day matter density.

Once the solution to this equation, $q(z)$ (where only the positive solution is the relevant one) has been obtained, it can be substituted in the expressions for $U'$ and thus of $H(z)$. $H(z)$, being the Universe expansion rate, is the key observable which can be obtained from astronomical observations.

\section{Observational Data and Analysis}

We constrain the $\lambda_i$ parameters  from the observations of the Hubble parameter as a function of redshift $H(z)$  in the redshift range $0.15 < z < 1.1$  from the cosmic chronometers project \cite{JL,Simon,stern,moresco}. In particular we use the latest determinations by Ref.~\cite{Moresco2012}.   To compare theory with observations we  use the Markov Chain Monte Carlo (MCMC) approach as anticipated in \cite{JTV11}. MCMC has become a standard workhorse in cosmology, thus we will not explain it in detail here, but a pedagogical introduction can be found e.g.,  in \cite{verdeproceedings}.
To explore the parameters space given by $\{H_0, \Omega_m, \lambda_0, \lambda_1,\lambda_2,\lambda_3,\lambda_4\}$ we use a standard Metropolis-Hasting algorithm, run 4 chains  starting from well separated points in parameter space and check convergence and mixing with the Gelman and Rubin \cite{gelmanrubin} criterion. We also impose flatness, and  the following priors: $H_0= 73.8\pm 2.4$ km s$^{-1}$ Mpc$^{-1}$ from \cite{Riessetal2011,moresco} and  $\Omega_m=0.27\pm 0.05$ from \cite{Komatsuwmap7}. 

\begin{figure}
\begin{center}
\includegraphics[width=.45\columnwidth]{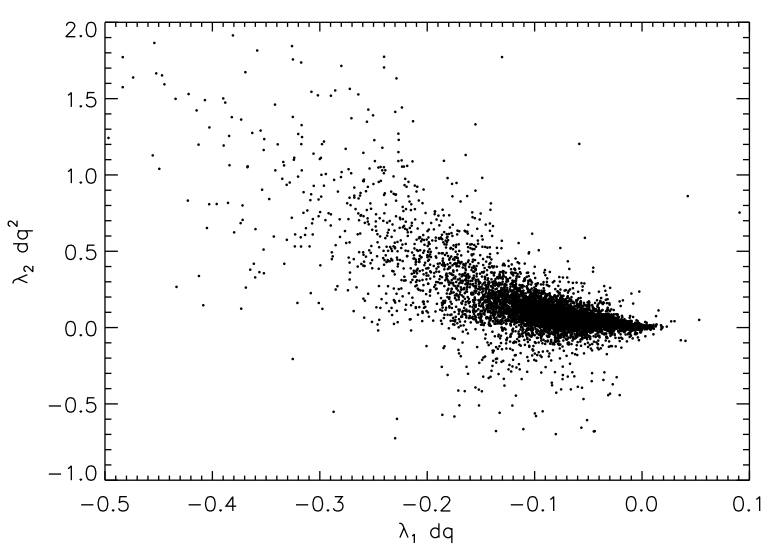}
\includegraphics[width=.45\columnwidth]{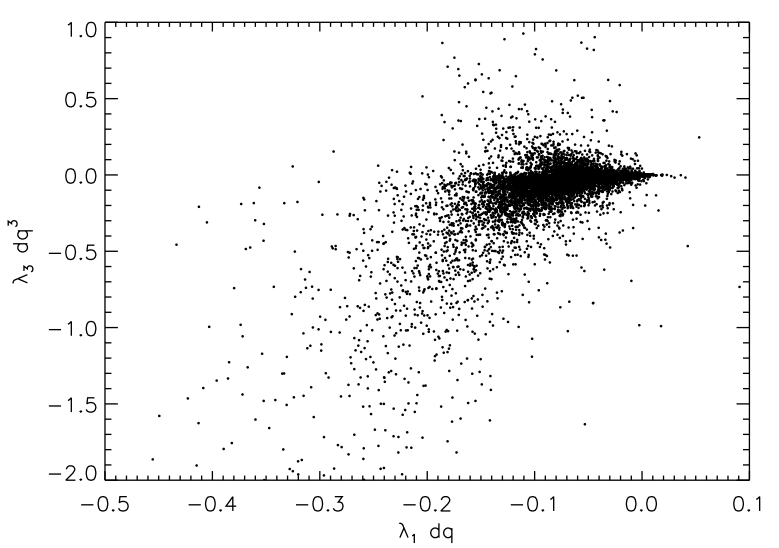}
\includegraphics[width=.45\columnwidth]{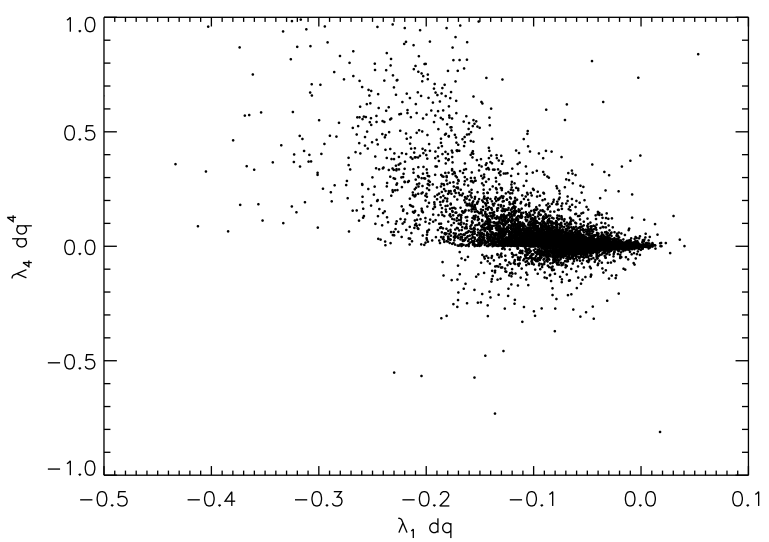}
\includegraphics[width=.45\columnwidth]{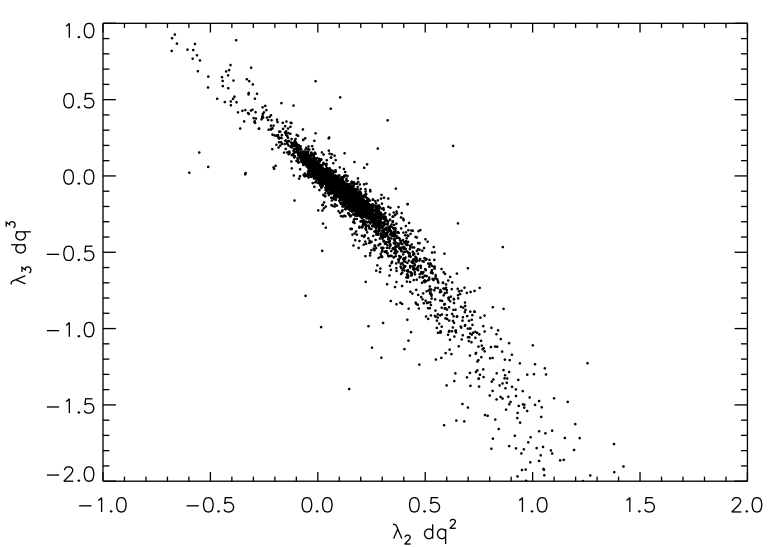}
\includegraphics[width=.45\columnwidth]{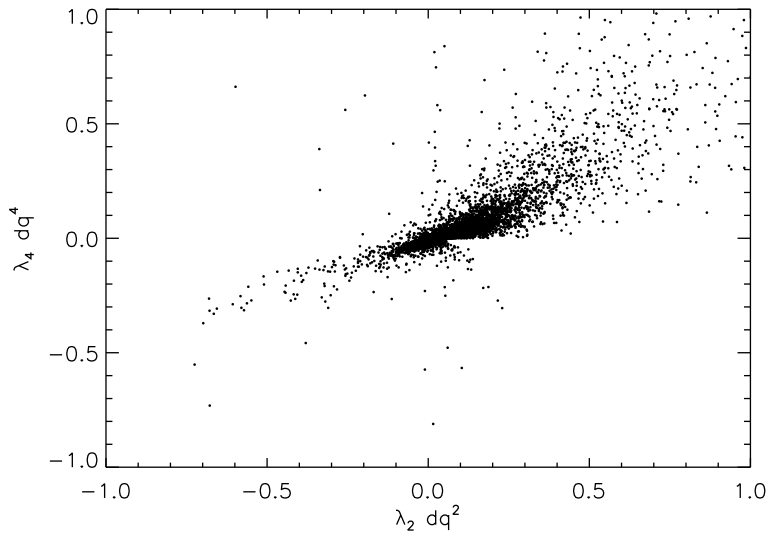}
\includegraphics[width=.45\columnwidth]{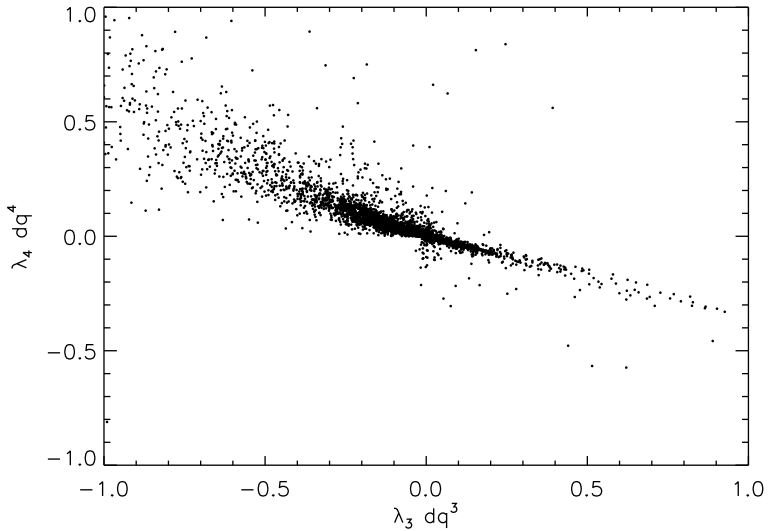}
\end{center}
\caption{Scatter plots for the different combinations of the free parameters in the potential ($\lambda_i$) weighted by the corresponding power in the field 
displacement. The points shown include many sigmas from the best fit value ($\sim 5 \sigma$). }
\label{fig:1}
\end{figure}

\begin{figure}
\begin{center}
\includegraphics[width=.45\columnwidth]{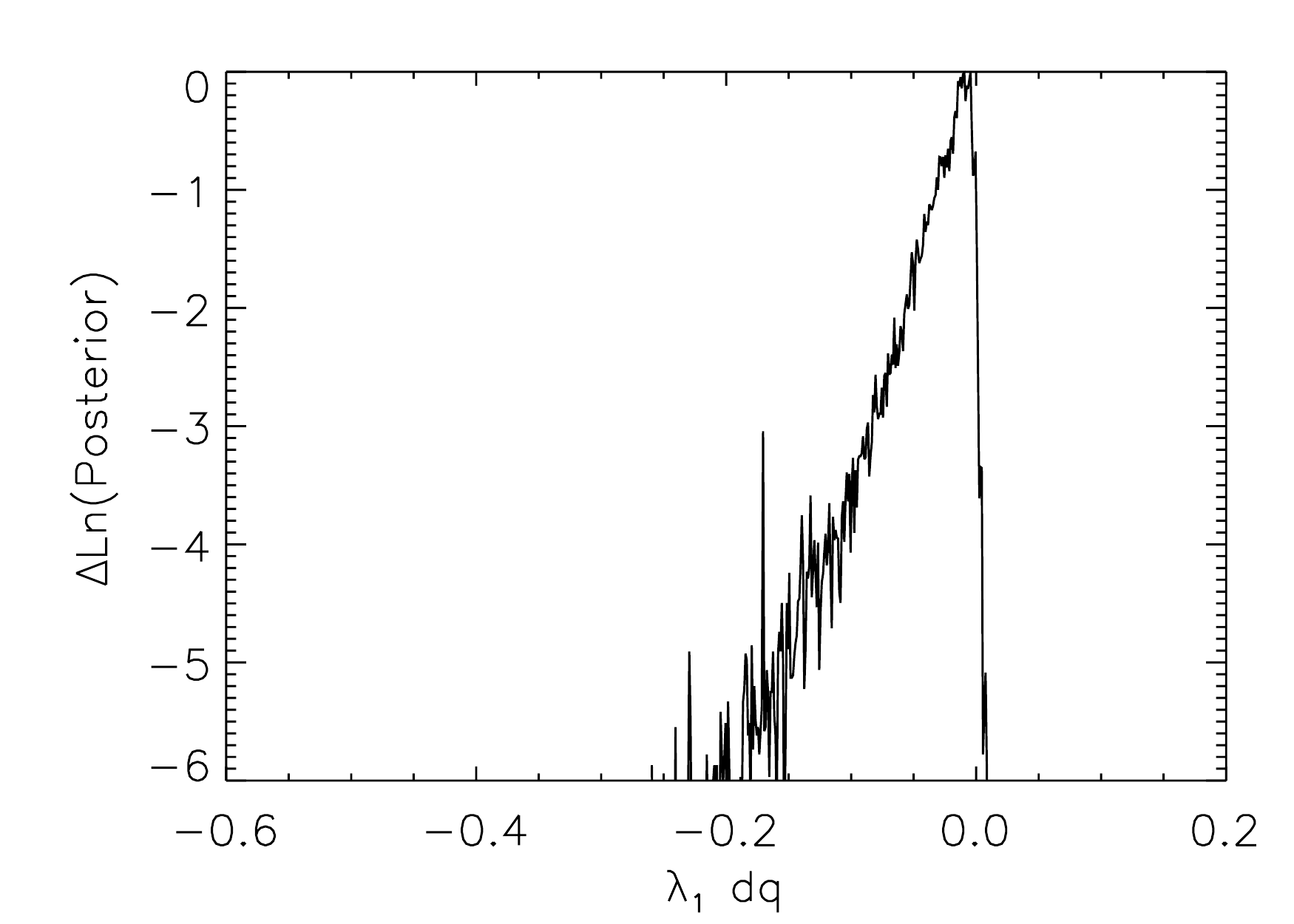}
\includegraphics[width=.45\columnwidth]{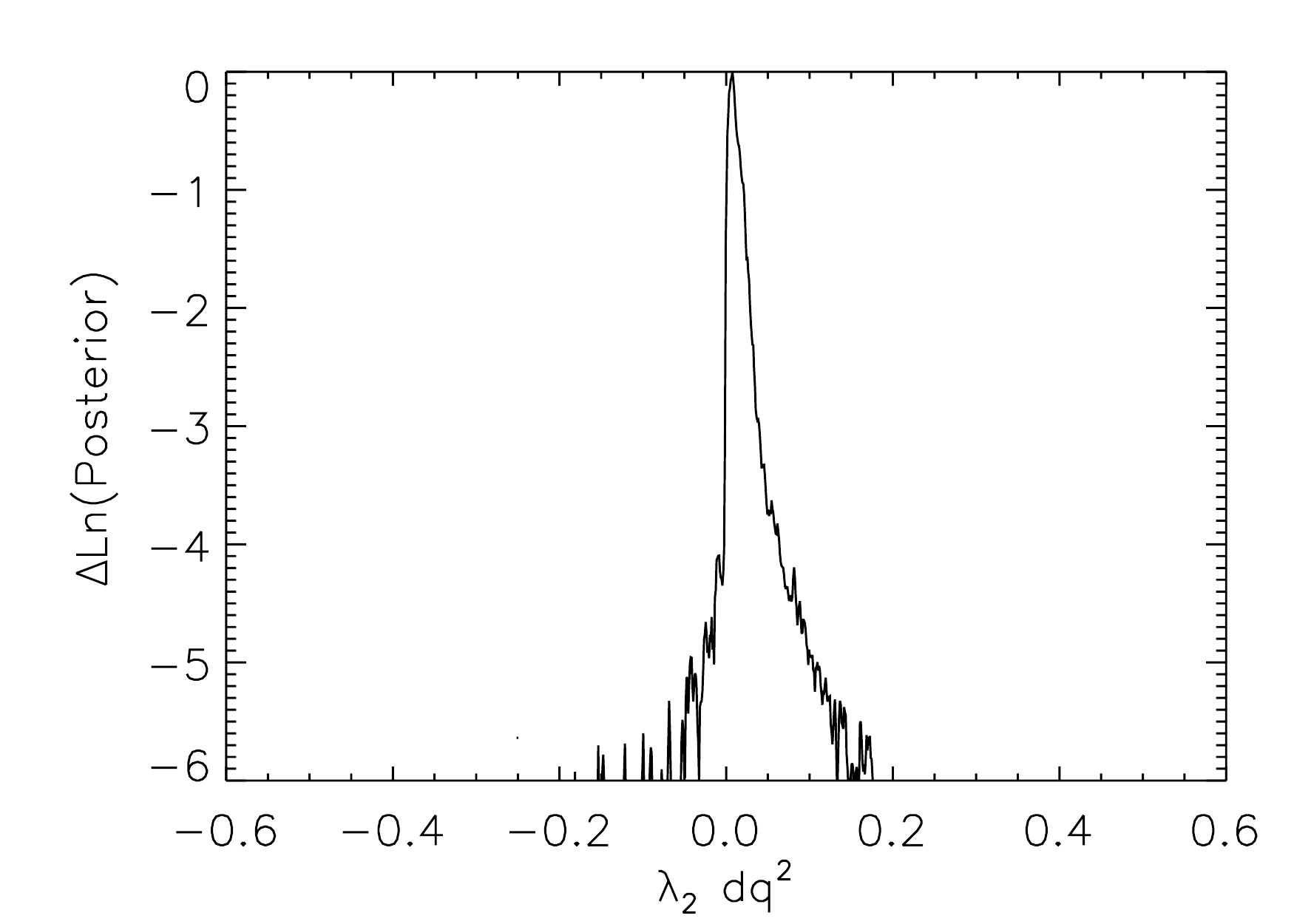}
\includegraphics[width=.45\columnwidth]{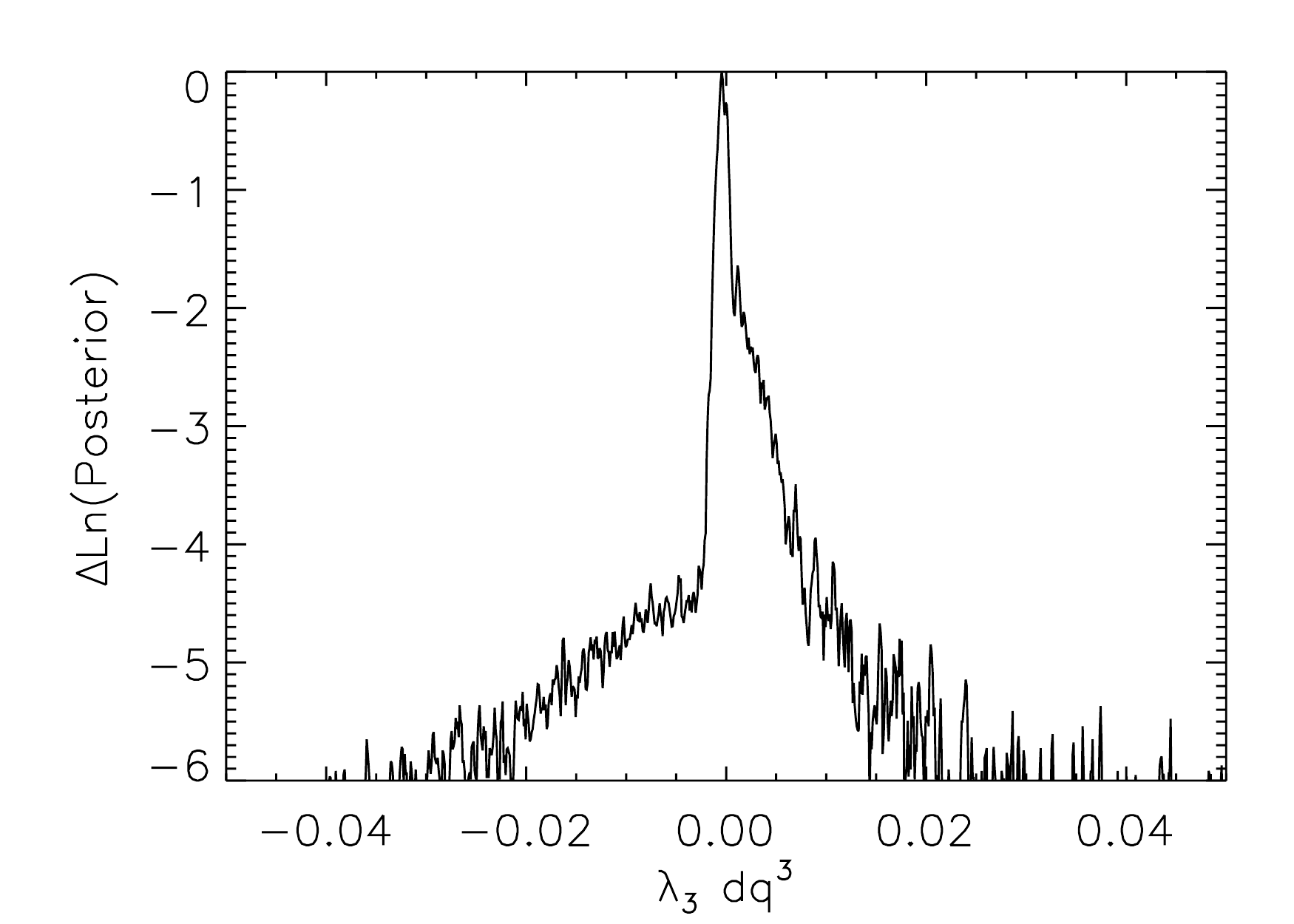}
\includegraphics[width=.45\columnwidth]{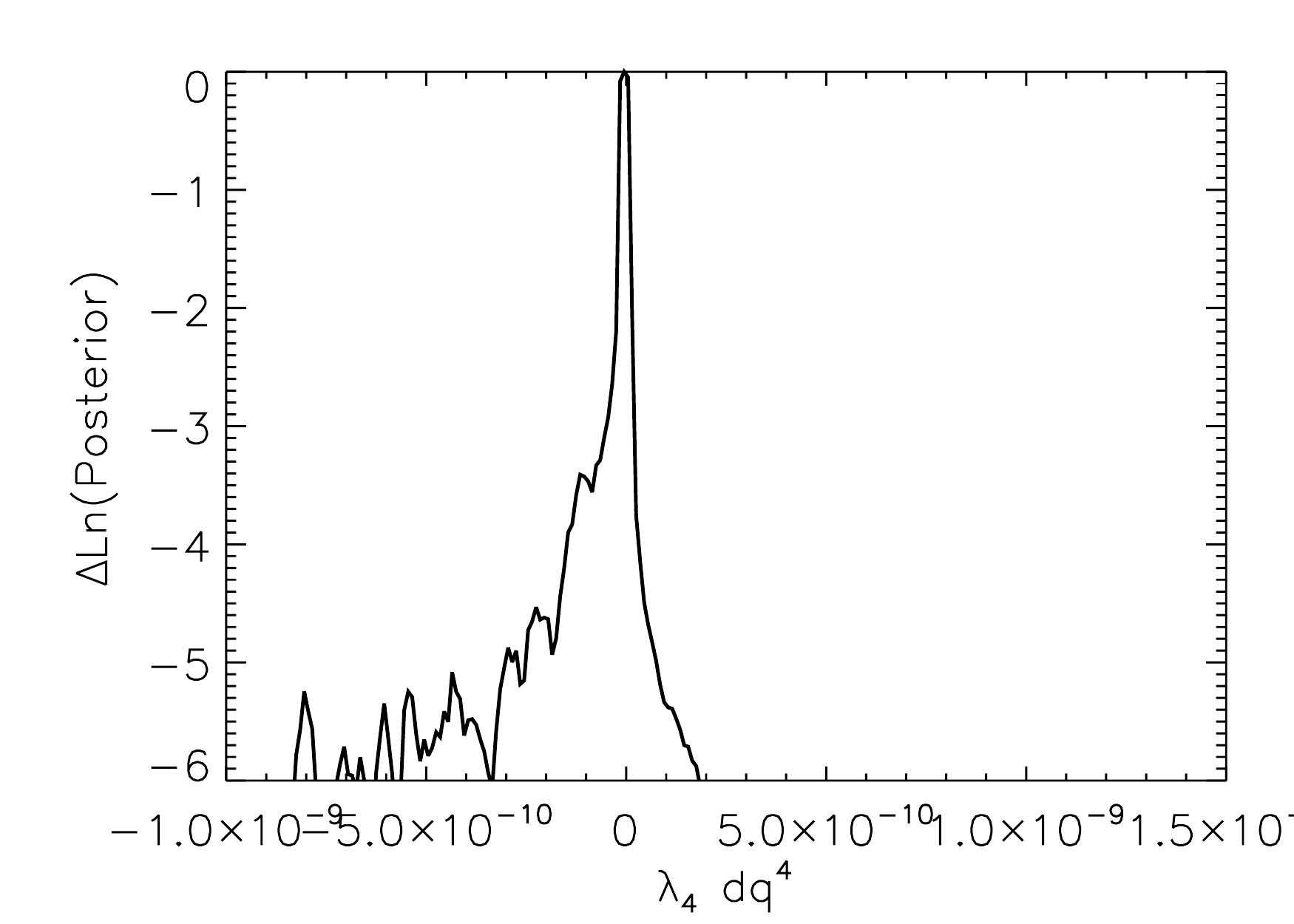}
\end{center}
\caption{1D posterior distribution for each of the parameters of the effective potential of accelerated expansion.}
\label{fig:2}
\end{figure}

\begin{figure}
\begin{center}
\includegraphics[width=.7\columnwidth]{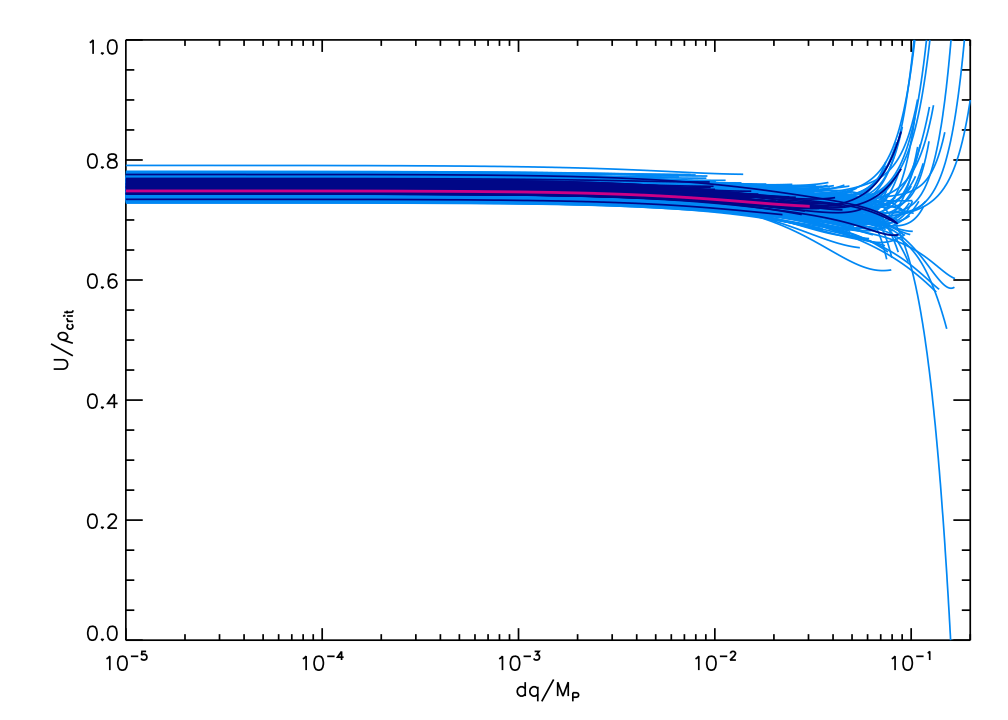}
\includegraphics[width=.7\columnwidth]{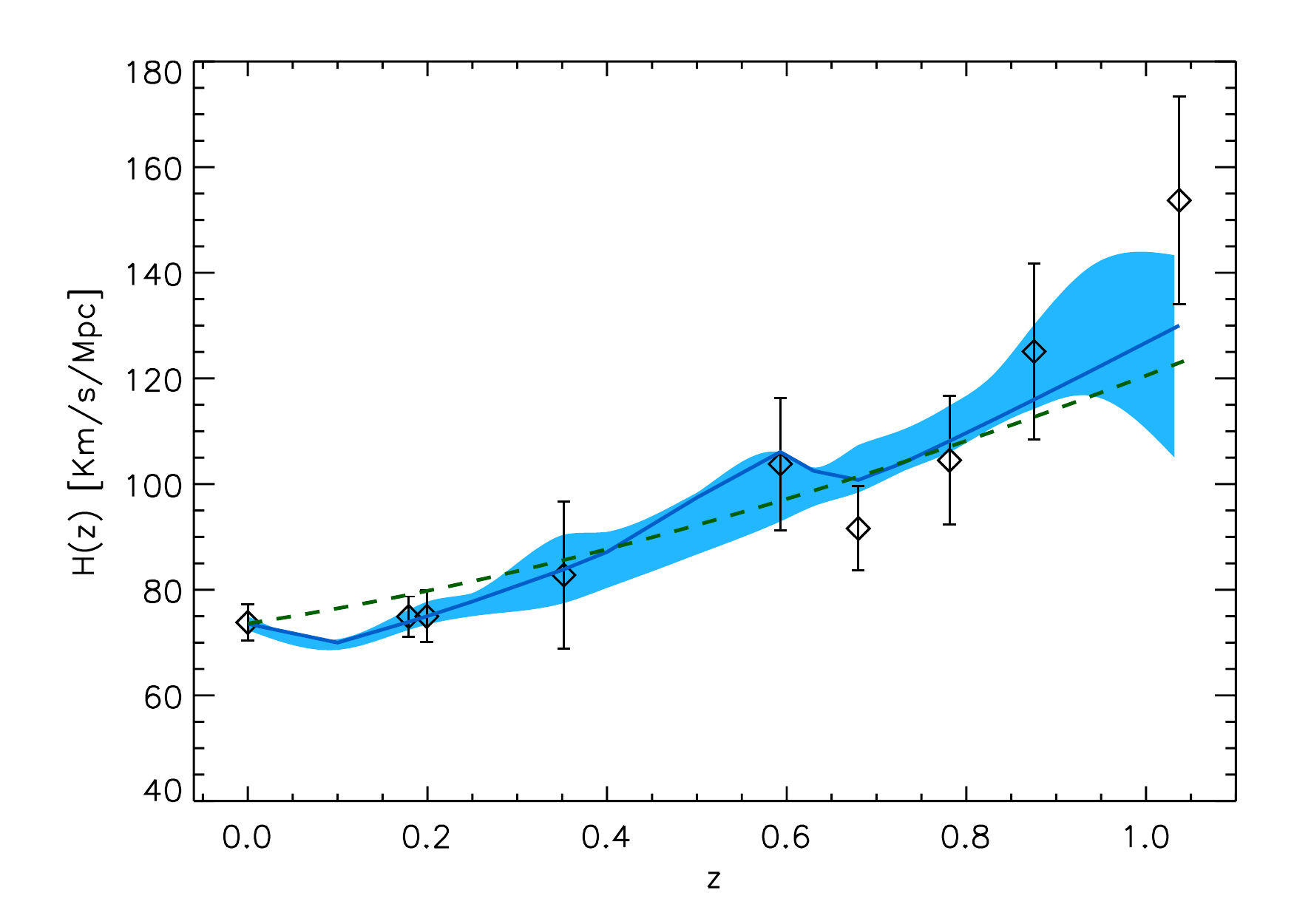}
\end{center}
\caption{Top panel: The effective potential of accelerated expansion $U(q)$ in units of the critical density $\rho_{\rm crit}$ as function of the displacement 
of the field $q$ in $M_{\rm p}$ units. Different tracks are plotted for values with 68\% confidence (dark blue lines) and 95\%  confidence (light blue lines). The best-fit model is shown as a solid red line, which is better described physically as a pseudo-Goldstone boson. The trajectories correspond to how much the field has moved in the full redshift range of the observational data ($0 < z < 1.1$).  Note that the potential is very flat at the few $6$\% level and that for many models the field displacement is very small. The data strongly favor a flat potential. Bottom panel: the $H(z)$ values used in this study and the best fitting models over-plotted as a blue region (68\% confidence). We also show the best fitting model as a light blue line and the LCDM model as a dashed black line.}
\label{fig:3}
\end{figure}

\section{Results}

The resulting  constraints on  the parameters  of the Lagrangian are shown in Fig.~\ref{fig:1}, where  we show the scatter plot of the models sampled by our MCMC. The results do not display any sign of the existence of discrete symmetries, as for instance $\varphi\to-\varphi$, that would signal the spontaneously breaking of a gauge symmetry \cite{Krauss:1988zc}.

In Fig.~\ref{fig:2} we show the corresponding 1-d marginalised posteriors on each of the $\lambda_i$. The $1\sigma$ confidence levels are: $0 < \lambda_1 < 2.57$, $\lambda_2 = 0.41_{-0.80}^{+1.00}$, $\lambda_3 =  0.07\pm 0.30$, $\lambda_4 =  0.02 \pm 0.02$.  These constraints will greatly improve  with forthcoming data which can in principle reduce the  errors on the reconstructed $H(z)$ by a factor of $3-5$. Notice that we have not imposed $\lambda_1=0$ as the field theory stability criterium implies. Despite of this, we can see from Fig.~\ref{fig:2} the the points in the MCMC with maximum likelihood are those with $\lambda_1=0$, which is a nice feature of the analysis. Although our  analysis points out so far the consistency of the cosmological constant picture, i.e. a flat potential $\lambda_i = 0$ $i\ge 1$, we cannot  rule out plausible alternatives as a dynamical vacuum or quintessence although such models would have to satisfy the constraints found here (Fig. 3 top panel). 

 In particular the result  yields a value of the scalar mass, $\lambda_2$, consistent with zero,  indicating  that  the scalar particle associated to the dark energy potential is consistent with a pseudo-Goldstone boson. We remark that,
 after constraining the theory with the data,
 the basic assumption of effective field theory seems to be verified, as the allowed ranges  for the $\lambda$ parameters  become smaller for higher-order $\lambda_i$. 

We show the reconstructed potential in Fig.~\ref{fig:3} -- which we have tested is insensitive to the choice of stellar population model (see Ref.~Moresco2012). The dark blue lines correspond to the $1\sigma$ while the light blue  lines to the $2\sigma$ confidence contours. The red line corresponds to the best fit to the data and the trajectories correspond to how much the field has moved in the full redshift range of the observational data ($0 < z < 1.1$). Note that the potential is very flat at the $6$\% level and that for many models the field displacement is very small. The data strongly favor a flat potential.

Finally let us consider the two relevant energy scales.

 {\sl i)} The first is the limit of  the range of validity of the effective theory. As an effective theory we expect to break down at some scale, $\mu$. We ignore which exact symmetry is broken by the pseudo-Goldstone boson and its mechanism,  but the reconstructed potential can give some insights. Fig.~\ref{fig:3}  shows that the potential deviates significantly from a constant  at  $\mu \approx 5\times 10^{-2} {\rm q\over m_p}$, that roughly coincides with the scale at which inflation ends, which is a curious finding. Notice that the slope can be either positive or negative signaling  that the cosmological constant can either decrease or increase as $z$ decreases. Whether the strong deviation at this point is driven by the error bars of the last experimental value or by a physical breaking of the perturbative approach is something to be  tested with better data at high redshift.

{\sl ii)} The second  energy scale is related to the breaking of unitarity. In order to set the scale of this effect, ${\cal M}$, we come back to (\ref{termsm6}) and calculate the
two--graviton scattering process.  We obtain that ${\cal M} \sim {\rm Min}\left[  {m_p\over \alpha_1} , {m_p^2\over \alpha_2}\right].$ As $\alpha_1\,,\alpha_2$
are treated as perturbation parameters one obtains that ${\cal M} \gg \mu$, thus the full approach breaks down before unitary is lost.

Notice that the scalar field acquires an effective mass due to its self-interaction. This scale settles the range of interaction, $m_{\rm eff}^{-1}$, of the {\sl fifth} force in the absence of matter
interaction 
\eqn{ef}{
m_{\rm eff}= {d^2U(q)\over dq^2}\vert_{q_c}= 2 \lambda_2 + 6 \lambda_3 q_c + 12 \lambda_4 q^2_c\,,  
}
where $q_c$ is the field value which minimizes the effective potential (\ref{poteff}). In our case the fifth force becomes negligible compared to gravity for a test object at a given distance. We will explore experimental constraint in a forthcoming publication.


\section{Discussion and Conclusions} 
We have confronted the  effective theory for accelerated expansion presented in \cite{JTV11} with data.  In particular we have  constrained the coefficient of the  leading terms in the effective Lagrangian  using observational data on the  Hubble parameter as a function of redshift  in the range $0<z<1.1$ from the cosmic chronometers project \cite{JL,Simon,stern,moresco,Moresco2012}.

 Our main finding  is that   the expansion history is consistent  with that predicted by a flat potential.  The data do not require extra parameters  beyond a constant term in the Lagrangian to explain the current accelerated expansion. Further, we have shown that the potential deviations from a constant are constrained to be below $6$\%.  Observational constraints allow the parameters describing the Lagrangian to vary only within certain limits; the  relative range of  the allowed variation of the  parameters confirms a  well defined hierarchy where the linear and quadratic terms dominate over higher-order terms, justifying the  basic assumption of the effective theory approach. Observational constraints also give some indications of the relevant energy scales involved.   Because a direct determination of a Lagrangian allows  us to determine the underlying symmetries in the theory, our results can be used to shed light on this as well. 

Our conclusions are model independent beside the assumption that dark energy is composed mainly by scalar particles. To go beyond this scenario  where dark energy is a minimally coupled scalar field, one needs to make extra assumptions about how dark energy interacts with other type of fields. Generically this is implemented by adding to (\ref{termsm6}) a Lagragian density of the type \cite{Carroll:1998zi}
\eqn{cou}{
\sum_i \alpha_i \varphi {\cal L}_i  
}
 where  $\alpha_i$ are  the coupling constants and ${\cal L}_i $ collects all gauge-invariant dimension-four operators.
For instance one possibility is to couple dark energy and neutrinos  
such that the neutrino masses are functions of the scalar field playing the role of dark energy \cite{astro-ph/0503349}. 
Another possibility is to interpret the light, but massive, scalar field as an axion-like particle that couples with the kinetic term of the Lagrangian of the photons \cite{arXiv:1007.0024}.
In  all these  cases the coupling constants $\alpha_i$ can be  directly inferred working within  our framework and using the most recent $H(z)$ data.

\section*{Acknowledgements}

The research of PT was supported in part by grant FPA2010-20807, grant 2009SGR502  and the grant Consolider CPAN.
LV is supported by FP7- IDEAS Phys.LSS 240117. LV and RJ are supported by MICINN grant AYA2008-03531.
MM and AC acknowledge financial contributions from contracts ASI-Uni Bologna-Astronomy Dept. Euclid- NIS I/039/10/0, and PRIN MIUR Dark energy and cosmology with large galaxy surveys. We thank Luis Alvarez-Gaume for useful discussions.

\end{document}